\documentclass[
]{ceurart}

\usepackage{listings}
\usepackage{booktabs}
\usepackage{tikz}
\usetikzlibrary{arrows.meta,positioning,fit,backgrounds}

\begin{document}

\copyrightyear{2026}
\copyrightclause{Copyright for this paper by its authors.
  Use permitted under Creative Commons License Attribution 4.0
  International (CC BY 4.0).}

\conference{Preprint. Not yet peer-reviewed.}

\title{Transaction Foundation Models for Retrieval-Augmented Generation
  over Financial Transaction Data: A Comparative Evaluation with Text
  Embeddings}

\author[1]{Snehal Pal}[%
email=palsnehal004@kgian.iitkgp.ac.in,
url=https://snehalpal.vercel.app/,
]
\cormark[1]
\address[1]{Indian Institute of Technology Kharagpur, Kharagpur, West Bengal, India}

\cortext[1]{Corresponding author.}

\begin{abstract}
  Domain-specific tabular and transaction foundation models are increasingly
  proposed as drop-in representation learners for financial applications, but
  their value relative to off-the-shelf text embedding models for
  retrieval-augmented generation (RAG) over structured data remains
  underexamined. We construct a deterministic, template-generated
  question-answering benchmark of 1{,}026 questions with retrieval-level
  gold-document annotations over IBM's synthetic Anti-Money-Laundering (AML)
  transaction dataset, and compare NVIDIA's 29M-parameter Transaction
  Foundation Model (TFM) against five general-purpose and financial-domain
  text embedding models under a size- and data-matched retrieval setup, as
  well as against a hand-aggregated summary-text baseline. TFM substantially
  outperforms every size-matched generic embedder on both retrieval
  (Recall@5: 0.81 vs.\ 0.25--0.39) and end-to-end answer accuracy (66.7\% vs.\
  27.5\%). However, a simple pipeline that embeds a human-engineered
  aggregated summary with a generic sentence encoder matches or exceeds TFM
  overall, and dominates it on several low-cardinality entity categories
  (bank, currency, payment format). This suggests that the benefit of a
  transaction-native foundation model over generic text embeddings is
  category-dependent rather than uniform, and that careful feature
  aggregation can substitute for domain-specific pretraining in several
  practically relevant regimes. We release the benchmark generation and
  evaluation code.
\end{abstract}


\maketitle

\section{Introduction}

Retrieval-augmented generation (RAG) systems are increasingly built over
structured, tabular enterprise data such as financial transaction logs,
where the standard practice is to linearize or aggregate rows into text and
embed the result with a general-purpose sentence encoder. An alternative,
recently popularized by domain-specific \emph{tabular} or \emph{transaction}
foundation models, is to represent structured rows directly with an encoder
pretrained on transaction sequences, avoiding a potentially lossy
text-conversion step.

This raises a practical question for anyone building a RAG system over
financial data: \emph{does retrieving with a domain-specific transaction
foundation model outperform retrieving with generic or financial-domain
text embedding models, and if so, under what conditions?}

We answer this empirically using NVIDIA's Transaction Foundation Model
(TFM)~\cite{wu2026transaction}, a 29M-parameter decoder pretrained on
transaction sequences, evaluated against five text embedding baselines on a
retrieval-augmented question-answering benchmark we construct over IBM's
synthetic AML transactions dataset~\cite{altman2023realistic}. Our
contributions are:

\begin{enumerate}
  \item A deterministic QA benchmark (1{,}026 questions, 161 retrieval
    documents) over structured financial transaction data, with every answer
    computed directly from the underlying data (no LLM in the ground-truth
    loop) and every question carrying gold document identifiers, enabling
    retrieval evaluation independent of generation evaluation.
  \item A controlled, data- and token-budget-matched comparison of TFM
    against five general and financial-domain text embedding models, plus a
    realistic human-engineered aggregated-summary baseline.
  \item A two-tier evaluation --- retrieval accuracy (Recall@K, MRR) and
    end-to-end answer accuracy via a shared generation model --- and a
    category-level analysis showing that TFM's advantage over generic text
    embeddings is concentrated in individual-entity, high-cardinality
    queries and largely absent (or reversed) for low-cardinality categorical
    queries.
\end{enumerate}

\textbf{Code and data.} The benchmark generation notebook and the
retrieval/evaluation notebook are both publicly available on Kaggle:
dataset generator\footnote{\url{https://www.kaggle.com/code/palsnehal004/rag-evaluation-dataset-generator-aml-transaction}}
and retriever evaluation\footnote{\url{https://www.kaggle.com/code/palsnehal004/tfm-vs-generic-embedding-models-rag-retriever-sh}}.

\section{Related Work}

Dense passage retrieval~\cite{karpukhin2020dense} and retrieval-augmented
generation~\cite{lewis2020retrieval} are now standard components of
knowledge-intensive NLP pipelines. General-purpose sentence embedding
models such as BGE~\cite{xiao2023cpack}, E5~\cite{wang2022text}, and
GTE~\cite{li2023towards} are commonly used as retrievers in such pipelines,
including over data that originates in structured or tabular form.
Financial-domain adaptations of encoder-only models, such as
FinBERT~\cite{araci2019finbert}, were built for sentiment classification
rather than retrieval, and are typically repurposed for embedding via
mean-pooling. Separately, tabular and transaction foundation models such as
TabFormer~\cite{padhi2021tabular} and NVIDIA's Transaction Foundation
Model~\cite{wu2026transaction} learn representations directly from
structured transaction sequences. To our knowledge, prior RAG evaluation
benchmarks for structured financial data rely on human-curated question
sets; our benchmark instead guarantees ground-truth correctness by
computing every answer directly from the source data via deterministic
aggregation.

\section{Dataset and Benchmark Construction}

We use the \emph{HI-Small} split of IBM's synthetic Anti-Money-Laundering
transactions dataset~\cite{altman2023realistic}: 5{,}078{,}345 transactions
across 515{,}088 unique accounts and 30{,}470 banks, with a 0.10\%
laundering rate.

We generate a retrieval corpus of 161 entity-level documents (one text
summary per unique account, bank, currency, payment format, date, sender-
receiver pair, or the whole dataset) and 1{,}026 question--answer pairs
from 35 templates across seven categories, sampling up to 40 entities per
template. Every answer is computed directly via pandas aggregation over the
raw transaction rows, guaranteeing ground truth without relying on an LLM.
Every QA item additionally carries a \texttt{gold\_doc\_ids} field linking
it to the supporting corpus document(s), which decouples retrieval
evaluation from generation evaluation. Table~\ref{tab:categories} shows the
category distribution.

\begin{table}
  \caption{QA pairs per question category}
  \label{tab:categories}
  \small
  \begin{tabular}{lr}
    \toprule
    Category & \# questions \\
    \midrule
    account   & 680 \\
    bank      & 160 \\
    pair      & 80 \\
    currency  & 45 \\
    date      & 36 \\
    format    & 21 \\
    global    & 4 \\
    \midrule
    Total     & 1{,}026 \\
    \bottomrule
  \end{tabular}
\end{table}

\section{Methodology}

\begin{figure}[t]
  \centering
  \includegraphics[width=0.8\linewidth]{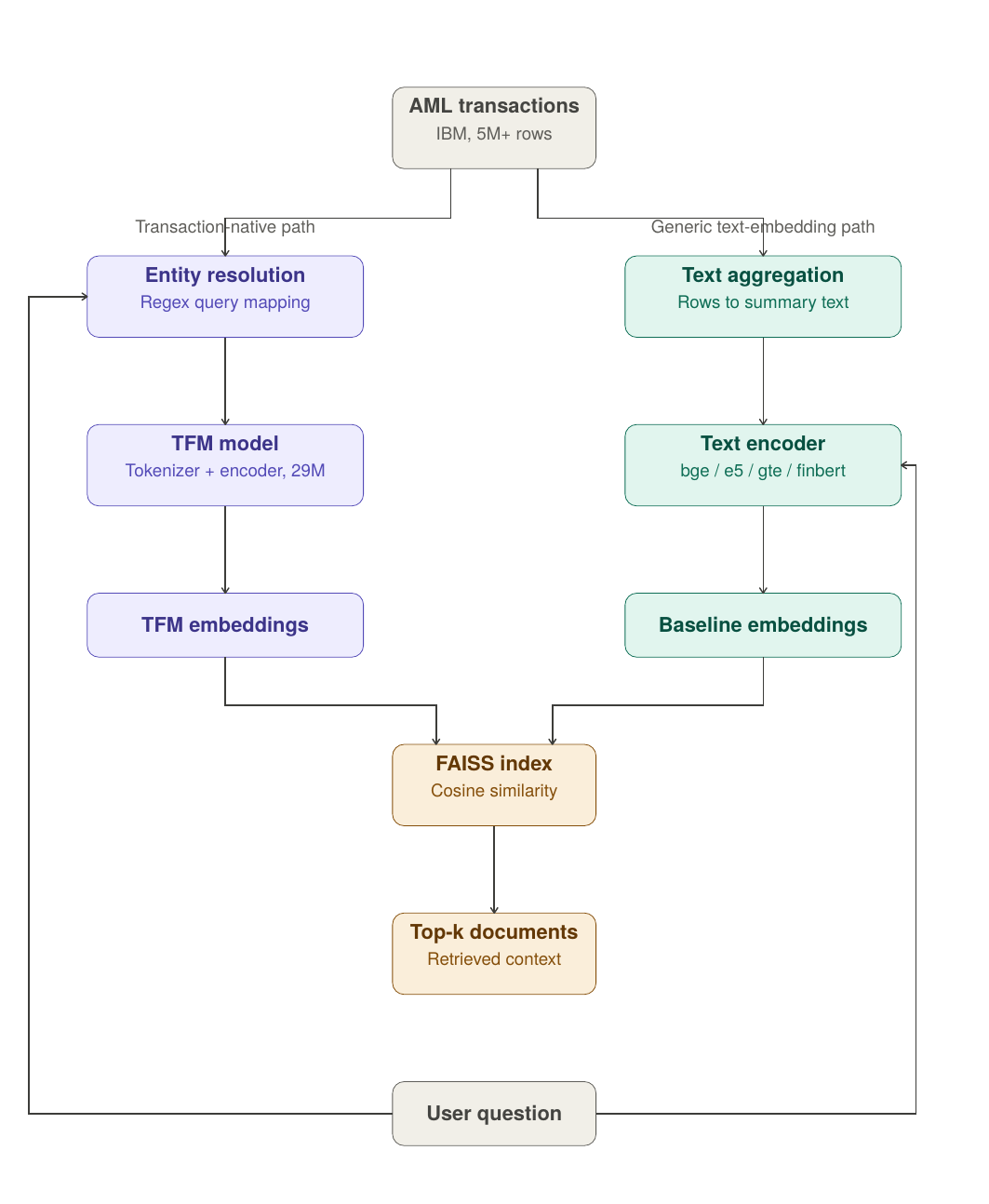}
  \caption{Retrieval representation pipeline. Identical raw transaction
    rows feed both the TFM path (entity resolution, the 29M-parameter TFM
    tokenizer and encoder, and TFM embeddings) and the text-embedding path
    (row aggregation into summary text, followed by a generic or
    financial-domain encoder: \texttt{bge-small}, \texttt{e5-small-v2},
    \texttt{gte-small}, \texttt{finbert}, or \texttt{finlang-investopedia}).
    Both representations are indexed identically with FAISS. The user
    question feeds \emph{both} the TFM path (via regex-based entity
    resolution) and the text-embedding path (embedded directly by the text
    encoder), so retrieval can be evaluated for either representation
    without any change to the query.}
  \label{fig:architecture-retrieval}
\end{figure}

\begin{figure}[t]
  \centering
  \includegraphics[width=\linewidth]{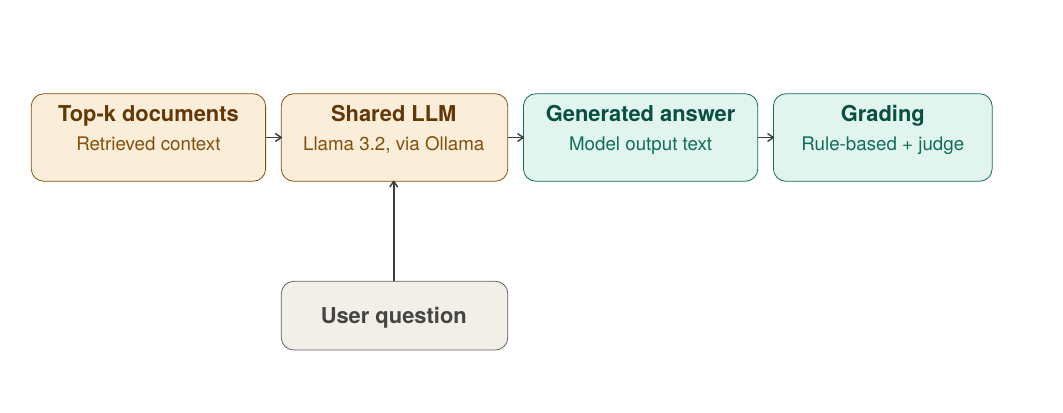}
  \caption{Generation and grading pipeline. The top-$k$ documents retrieved
    in Figure~\ref{fig:architecture-retrieval} and the original user
    question are passed to the same shared generation model (Llama 3.2 1B
    via Ollama) regardless of which retriever produced the retrieved
    documents, so only the retrieval representation varies across
    conditions. Generated answers are graded against the gold answer via
    rule-based matching with an optional LLM-judge pass.}
  \label{fig:architecture-generation}
\end{figure}

Figures~\ref{fig:architecture-retrieval} and~\ref{fig:architecture-generation}
summarize the overall pipeline: raw AML transaction rows are represented
two ways --- through TFM's native tokenizer and encoder, and through text
aggregation followed by a text-embedding model --- both indexed with the
same FAISS search structure and evaluated with the same retrieval metrics
and the same downstream generation model.

\subsection{Retrievers under test}

\textbf{A: TFM (domain).} NVIDIA's 29M-parameter decoder-only Transaction
Foundation Model~\cite{wu2026transaction}, pretrained on card-transaction
sequences following the TabFormer field schema. AML fields are mapped onto
TFM's expected schema (amount bucket from \emph{Amount Paid}; a hash of the
counterparty account as the merchant field; hour, day-of-week, and month
from \emph{Timestamp}; \emph{Payment Format} as the card-type field); fields
absent from AML data receive fixed defaults. Each entity's raw transaction
rows are tokenized under this schema and embedded via last-token pooling.

\textbf{B (fair) baselines.} Five general and financial-domain sentence
embedding models --- \texttt{bge-small}, \texttt{e5-small-v2},
\texttt{gte-small}, \texttt{finbert}, and \texttt{finlang-investopedia} ---
each embed a text paragraph built from the \emph{identical} raw transaction
rows TFM consumes for the same entity, under the same 512-token budget.
This isolates the effect of the representation model while holding the
underlying data fixed. FinBERT embeddings use mean-pooling over the base
encoder's last hidden state, since it is a classification model with no
native sentence-embedding head.

\textbf{B1 (legacy).} \texttt{bge-small} embedding the pre-aggregated,
human-engineered summary text used to build the retrieval corpus itself
(e.g., ``Account X (bank Y) summary: sent N transactions totaling \ldots'')
rather than raw transaction rows. This represents a realistic, already
feature-engineered text pipeline, as opposed to a naive one.

\subsection{Query-side resolution}

TFM cannot embed natural-language questions directly. We resolve each
question to its target entity via category-specific regular expressions
over the fixed template phrasings (verified to match all cases across all
categories in the released QA set), and embed that entity's held-out
transaction slice --- rows excluded from the indexed slice --- through TFM
as the query representation, so retrieval is not a trivial self-match. Text
embedding baselines embed the question text directly.

\subsection{Indexing, retrieval, and generation}

All document embeddings are L2-normalized and indexed with a FAISS flat
inner-product index (equivalent to cosine similarity). We report
Recall@\{1,3,5,10\} and Mean Reciprocal Rank (MRR) against
\texttt{gold\_doc\_ids}. For end-to-end evaluation, the top-$k$ retrieved
documents are concatenated as context and passed, together with the
question, to a single shared local LLM (Llama 3.2 1B via Ollama,
temperature 0, JSON-constrained single-value output), held constant across
retrievers so only the retrieved context varies. Generated answers are
graded with exact/tolerant rule-based matching (numeric answers within a
relative tolerance, ISO dates, and categorical substring match), with an
optional secondary-model judge pass restricted to rule-based misses, using
a model distinct from the generator to avoid self-grading leniency bias.

\section{Experimental Setup}

All 1{,}026 QA pairs are evaluated; the retrieval corpus is never
subsampled, so every retriever always searches the full 161-document pool.
We report Recall@$K$ for $K \in \{1,3,5,10\}$ and MRR, with a fixed random
seed (42) throughout.

\section{Results}

Table~\ref{tab:overall} reports overall retrieval and end-to-end answer
accuracy. Table~\ref{tab:category} breaks down Recall@5 and MRR by question
category for TFM, the B1 legacy baseline, and the strongest fair generic
baseline (\texttt{gte-small}).

\begin{table}
  \caption{Overall retrieval metrics and end-to-end answer accuracy
    (accuracy reported only for retrievers with a completed generation
    pass)}
  \label{tab:overall}
  \footnotesize
  \setlength{\tabcolsep}{4pt}
  \begin{tabular}{lrrrrrr}
    \toprule
    Retriever & R@1 & R@3 & R@5 & R@10 & MRR & Acc.\ (judged) \\
    \midrule
    A: TFM (domain)                   & 0.758 & 0.787 & 0.805 & 0.846 & 0.780 & 0.672 \\
    B1: bge-small on summary (legacy) & 0.698 & 0.786 & 0.823 & 0.869 & 0.752 & 0.662 \\
    bge-small (fair)                  & 0.132 & 0.210 & 0.252 & 0.320 & 0.185 & 0.280 \\
    e5-small-v2 (fair)                & 0.136 & 0.243 & 0.306 & 0.396 & 0.209 & -- \\
    gte-small (fair)                  & 0.255 & 0.346 & 0.391 & 0.475 & 0.317 & -- \\
    finbert (fair)                    & 0.062 & 0.103 & 0.135 & 0.209 & 0.095 & -- \\
    finlang-investopedia (fair)       & 0.009 & 0.025 & 0.039 & 0.079 & 0.025 & -- \\
    \bottomrule
  \end{tabular}
\end{table}

\begin{table}
  \caption{Recall@5 / MRR by question category}
  \label{tab:category}
  \footnotesize
  \setlength{\tabcolsep}{4pt}
  \begin{tabular}{lrrrrrr}
    \toprule
    & \multicolumn{2}{c}{TFM (domain)} & \multicolumn{2}{c}{B1 (legacy)} & \multicolumn{2}{c}{gte-small (fair)} \\
    Category & R@5 & MRR & R@5 & MRR & R@5 & MRR \\
    \midrule
    account  & 0.825 & 0.829 & 0.751 & 0.656 & 0.276 & 0.217 \\
    bank     & 0.875 & 0.863 & 1.000 & 0.981 & 0.694 & 0.605 \\
    currency & 0.600 & 0.239 & 1.000 & 1.000 & 0.711 & 0.444 \\
    date     & 0.722 & 0.561 & 0.639 & 0.511 & 0.389 & 0.328 \\
    format   & 0.571 & 0.413 & 1.000 & 0.976 & 0.048 & 0.028 \\
    pair     & 0.700 & 0.693 & 1.000 & 1.000 & 0.688 & 0.608 \\
    global   & 1.000 & 1.000 & 1.000 & 1.000 & 0.000 & 0.000 \\
    \bottomrule
  \end{tabular}
\end{table}

\section{Discussion}

TFM's advantage is largest on the \emph{account} category, which dominates
the benchmark (680/1{,}026 questions): it beats the B1 legacy baseline by
7.4 recall points and every fair generic embedder by a wide margin,
consistent with its pretraining on individual transaction sequences
capturing fine-grained per-account behavior that a single aggregated
summary sentence compresses away.

On low-cardinality categorical questions --- bank, currency, payment
format, sender--receiver pair --- the picture reverses. B1's single
aggregated-summary embedding achieves near-perfect Recall@5 (0.94--1.00
across these four categories), while TFM trails (0.57--0.88) despite being
the domain-specific model. With only a handful of distinct banks,
currencies, or payment formats relative to the number of accounts, a short
lexical or semantic summary discriminates almost perfectly; TFM's
finer-grained transaction-sequence representation offers no advantage here
and, for currency and format, actively underperforms even the fair generic
\texttt{gte-small} baseline on MRR.

At the end-to-end level, TFM (67.2\% judged accuracy) narrowly beats B1
(66.2\%), both far ahead of the fair generic embedder tested end-to-end
(bge-small, 28.0\%). Because \emph{account} questions make up two-thirds of
the benchmark, TFM's per-account retrieval edge is enough to give it the
overall lead in end-to-end accuracy despite losing to B1 on retrieval
Recall@5 overall (0.805 vs.\ 0.823) and on four of seven categories.

The overall picture is that the value of a transaction-native foundation
model over generic text embeddings is \emph{category-dependent}: it earns
its keep specifically for individual-entity, high-cardinality behavioral
queries, while a well-engineered generic-embedding pipeline over aggregated
summary text is a strong, considerably cheaper substitute for coarse-grained
categorical lookups.

\section{Limitations}

Our benchmark is built on a single synthetic dataset and domain (AML
transactions); questions are generated from 35 fixed templates and so lack
the linguistic variety of real user queries; end-to-end accuracy is
measured with a single, small (1B-parameter) generation model; results are
reported from a single seed without confidence intervals, which matters
most for the smallest categories (\texttt{global}, $n=4$); and among the
fair baselines, \texttt{finlang-investopedia} has roughly three times TFM's
parameter count, so it is not a size-matched comparison.

\section{Conclusion and Future Work}

We introduced a deterministic RAG evaluation benchmark over financial
transaction data and used it to show that a domain-specific transaction
foundation model outperforms size-matched generic text embedding models
overall, but that this advantage is concentrated in individual-entity
queries and disappears or reverses for low-cardinality categorical queries,
where a well-engineered generic-embedding pipeline is competitive or
better. Future work includes evaluating on additional domains and
datasets, testing paraphrased and free-form queries, repeating the
end-to-end evaluation with larger generation models, and reporting
variance across multiple seeds.

\section*{Declaration on Generative AI}
During the preparation of this work, the author used a generative AI
assistant (Anthropic's Claude) to help draft and edit the manuscript text
and to organize experimental results already produced by the author's own
notebooks into the tables and analysis presented above. All experimental
design, code, data generation, and result computation were performed by
the author; the AI assistant was not used to generate, alter, or
fabricate any data, results, or citations. The author reviewed and edited
all AI-assisted text and takes full responsibility for the content of this
publication.

\bibliography{refs}

\end{document}